\documentclass[reprint,
superscriptaddress,
amsmath,amssymb,
aps,
prc,
floatfix,
]{revtex4-2}

\usepackage[T1]{fontenc}
\usepackage{lmodern}

\usepackage{siunitx}
\usepackage{graphicx}
\usepackage{dcolumn}
\usepackage{bm}
\usepackage{ulem}
\usepackage{multirow}
\usepackage{comment}
\usepackage{hyperref}
\usepackage{colortbl}
\usepackage{xcolor}

\begin{document}

\title{Mass and decay width of $T_{ccs}$ from symmetries}

\author{Mitsuru Tanaka}
\email{tanaka@hken.phys.nagoya-u.ac.jp}
\affiliation{Department of Physics, Nagoya University, Nagoya 464-8602, Japan}
\author{Yasuhiro Yamaguchi}
\email{yamaguchi@hken.phys.nagoya-u.ac.jp}
\affiliation{Department of Physics, Nagoya University, Nagoya 464-8602, Japan}
\affiliation{Kobayashi-Maskawa Institute for the Origin of Particles and the Universe, Nagoya University, Nagoya, 464-8602, Japan}
\affiliation{Meson Science Laboratory, Cluster for Pioneering Research, RIKEN, Hirosawa, Wako, Saitama 351-0198, Japan}
\author{Masayasu Harada}
\email{harada@hken.phys.nagoya-u.ac.jp}
\affiliation{Kobayashi-Maskawa Institute for the Origin of Particles and the Universe, Nagoya University, Nagoya, 464-8602, Japan}
\affiliation{Department of Physics, Nagoya University, Nagoya 464-8602, Japan}
\affiliation{Advanced Science Research Center, Japan Atomic Energy Agency, Tokai 319-1195, Japan}

\date{\today}

\begin{abstract}
We analyze the mass and width of the doubly heavy tetraquark $T_{ccs}$ composed of a heavy diquark and a light-quark cloud with strangeness with assuming that a color antitriplet heavy diquark is a dominant component of the doubly charmed tetraquarks $T_{cc}$ and $T_{ccs}$. 
We construct an effective Lagrangian for masses of heavy hadrons based on the superflavor symmetry between the doubly heavy tetraquarks and the singly heavy baryons by including the terms that simultaneously break the heavy-quark and light-flavor symmetries, and predict the mass of $T_{ccs}$ as $M(T_{ccs}) = 4047\pm11$\,MeV.
The comparison of this prediction with future experimental observation will give a clue to understand the color structure of the heavy diquark.
We also predict the mass of $\Omega_{cc}$ as $M(\Omega_{cc}) = 3706^{+14}_{-15}\,$MeV.
We next calculate the decay width of $T_{ccs}$, based on solely the light-flavor symmetry, as $\Gamma(T_{ccs}) = 42\pm 24$\,MeV.
\begin{description}
\item[Keywords]
superflavor symmetry, heavy quark symmetry, flavor symmetry, diquark, doubly heavy tetraquark
\end{description}
\end{abstract}

\maketitle

\section{\label{sec:level1}INTRODUCTION}
The discovery of $X(3872)$~\cite{Belle:2003nnu} 
marked the beginning of numerous exotic hadron discoveries in the heavy-quark sectors, yet their structure remains poorly understood.
Hadrons with exotic structures beyond ordinary baryons ($qqq$) and  mesons ($q\bar{q}$) were already indicated by Gell-Mann and Zweig in the 1960s~\cite{Gell-Mann:1964ewy,Zweig:1964ruk,Zweig:1964jf,Swanson:2006st,Yamaguchi:2019vea}. 
Possible structures of the multiquark state being a color singlet have been discussed in the literatures (See, for reviews, e.g.~Refs.~\cite{Chen:2022asf,Meng:2022ozq}.)
The compact multiquark has been investigated as a color singlet state of few-body multiquark systems by the constituent quark model, etc. (e.g., Refs.~\cite{Vijande:2003ki,PhysRevD.95.014010,Hiyama:2018ukv,Meng:2019fan,Zhang:2020cdi,Wang:2022yes,Yang:2023dzb,Huang:2023jec}.)
The emergence of the hadronic molecules as a deuteronlike state, discussed as a deuson in Ref.~\cite{Tornqvist:1993ng}, is expected near the thresholds.
In fact many candidates of a hadronic molecule have been reported in the experimental studies as the $XYZ$ tetraquarks being a meson-meson state and the $P_c$ pentaquarks being a meson-baryon one.
Investigating the exotic structures would lead to an understanding of the QCD phenomena such as color confinement.

The doubly charmed tetraquark $T_{cc}^{+}$ was reported in the LHCb experiment in 2021 \cite{LHCb:2021vvq,LHCb:2021auc}. 
The reported state is consistent with a genuine exotic hadron having a flavor structure $cc\bar{u}\bar{d}$.
The spin and parity of $T_{cc}^{+}$ are determined to be $J^{P}=1^{+}$, and the LHCb considers $T_{cc}^{+}$ as an isoscalar.
The mass of $T_{cc}^+$ is $\SI{3874.817}{MeV}$ close to the $D^0D^{\ast +}$ threshold.
The decay to $D^{0}D^{0}\pi^{+}$ has been confirmed, with a decay width of $\SI{410}{keV}$~\cite{LHCb:2021vvq} or $48\,$keV~\cite{LHCb:2021auc}.
Furthermore, the LHCb analysis supports that $T_{cc}^+$ decays to $D^{0}D^{0}\pi^{+}$ via the intermediate state $D^{*+}$.

Since the discovery of $T_{cc}^{+}$, research on the doubly heavy tetraquarks (DHTs) has been actively conducted~\cite{Chen:2022asf}.
However at present, no clear answer has been obtained regarding the structure of DHTs.
For example, analyses based on the hadronic molecular model, which assumes that $T_{cc}^{+}$ is a loosely bound state with $D$ and $D^{*}$, have been conducted~\cite{Manohar:1992nd,Janc:2004qn,Ohkoda:2012hv,Li:2012ss,Cheng:2022qcm,Wang:2021yld,Wang:2021ajy,Ren:2021dsi,Xin:2021wcr,Wang:2022jop,Xin:2022bzt,Asanuma:2023atv,Sakai:2023syt}.
This is due to the fact that the mass of $T_{cc}^{+}$ resides in the vicinity of the $D D^\ast$
threshold.
On the other hand, a compact tetraquark structure of DHT is considered~\cite{Kim:2022mpa}, based on the diquark picture proposed by Jaffe~\cite{Jaffe:1976ig}.
In addition, DHTs have 
been discussed in various approaches such as the string model~\cite{Andreev:2021eyj}, the QCD sum rules~\cite{Navarra:2007yw,Du:2012wp,Agaev:2021vur}, and the lattice QCD~\cite{Ikeda:2013vwa,Junnarkar:2018twb,Padmanath:2022cvl,Lyu:2023xro}.

Symmetries such as flavor symmetry and chiral symmetry play a crucial role in the classification of hadronic states.
For exotic hadrons including heavy-quarks, symmetries that emerge in the heavy quark limit, such as heavy-quark symmetry (HQS)~(see, e.g., Ref.~\cite{Manohar:2000dt}) and superflavor symmetry~\cite{Georgi:1990ak,Savage:1990di,Fleming:2005pd}, are considered potentially useful for understanding the structures of exotics.
The superflavor symmetry emerges in an exchange of an antiheavy quark and a heavy diquark with the same color configuration $\bar{\mathbf{3}}$ in the heavy-quark limit.
Because of the sufficiently large mass of the heavy diquark, the spin‒dependent color
magnetic force is negligible, and does not contribute in the heavy-quark limit.
In this context, the heavy diquark and an antiheavy quark behave as the static color $\bar{3}$ source and are equivalent in terms of color interaction.
The property of hadrons remains invariant under the interchange of the heavy diquark and antiheavy quark.
We will refer to hadrons related under this superflavor symmetry as superflavor partners.
For instance, an anti-heavy meson(HM) $\bar{Q}q$ and a doubly heavy  baryon(DHB) $QQq$ are superflavor partners.
Another color representation for diquarks, $6$, is also allowed in DHTs but not in ordinary hadrons.
If diquarks take the color $6$ representation, the superflavor symmetry does not arise.
Hence, superflavor symmetry might shed light on the color configuration of exotic hadrons.
We think that understanding the color configuration of the diquark in DHTs is important because the interaction changes according to the color representation.

To analyze the mass spectrum of DHTs in terms of superflavor symmetry, this study assumes that DHTs consist of a color $\bar{3}$ heavy diquark, treated as a spatially compact object, and a light-quark cloud surrounding it.
Although it may exist in a mixed state within a DHT, we also assume that the color antitriplet state is the dominant state of the diquark because the color $\bar{3}$ diquark is likely realized in the ground state~\cite{Carlson:1987hh,Meng:2020knc,Chen:2022asf}.
Here, ``being spatially compact''
means that the heavy diquark can be approximated as a pointlike particle, with no radial excitation occurring between two heavy quarks.
The analyses of $T_{bb}$, being the bottom counterpart of $T_{cc}$, based on the quark model~\cite{Meng:2020knc,Meng:2021yjr,He:2023ucd} suggest that the distance between the two bottom quarks is shorter compared to other quark distances.
This observation appears to support the notion that $T_{bb}$ is composed of a heavy diquark and a light-quark cloud.
In the present analysis, we assume that $T_{cc}$ also holds such heavy diquark structure.
For simplicity, this analysis focuses solely on diquarks where both heavy quarks are of the same flavor.
Under these conditions, a color $\bar{3}$ diquark possesses spin one ($S_{QQ}=1$), whereas a color $6$ diquark has spin zero ($S_{QQ}=0$).

If $T_{cc}^{+}$ is the superflavor partner of $\Lambda_c^{+}$, we can naturally expect the existence of $T_{ccs}$ as the superflavor partner of $\Xi_c$, which belongs to the same flavor multiplet as $\Lambda_c^{+}$.
We consider that investigation of $T_{ccs}$ is useful to understand the nature not only of $T_{ccs}$ itself but also of $T_{cc}^{+}$.
We analyze $T_{ccs}$ by using the experimental result of $T_{cc}^{+}$ as an input.
Thus if the results obtained in this paper are eventually consistent with results in future experiments, it can be interpreted that the color antitriplet state is dominant in DHTs $T_{cc}^+$ and $T_{ccs}$.
In the following, we first derive simple mass relations assuming that the heavy diquark is spatially compact and color antitriplet state together with superflavor symmetry.
We note that the relations agree with the ones derived in Ref.~\cite{PhysRevLett.119.202002}.
However, we find a discrepancy between the prediction by the simple mass relation and the recent experimental data of $T_{cc}^{+}$.
Thus, we invent the improved mass relations with correction terms violating the symmetries and a mixing term of color $\bar{3}$ and $6$ states of the $cc$ diquark.
Then, we obtain new relations among HMs, DHBs, singly heavy baryons (SHBs) and DHTs, and predict the mass of $T_{ccs}$.
Furthermore, we also predict the decay width of $T_{ccs}$ from the $\rm{SU(3)}$ flavor symmetry. 

This paper is organized as follows. In Sec,~\ref{simplerelation}, we derive the simple mass relations among singly heavy and doubly heavy hadrons based on the heavy quark and superflavor symmetries.
We point out the existence of a discrepancy between the theoretical formulas and the experimental results.
In Sec.~\ref{sec:IMPROVED-MASS-RELATION}, we construct the effective Lagrangians including corrections and obtain the improved mass relations.
The mass of $T_{ccs}$ and in addition, the one of $\Omega_{cc}$ are predicted.
In Sec.~\ref{width}, the decay width of $T_{ccs}$ is predicted by using the effective Lagrangian approach respecting the flavor symmetry.
Finally, Sec.~\ref{sec:summary} is devoted to the summary.

\section{
Simple mass relations from superflavor and light-flavor symmetries
} \label{simplerelation}

In this section, we first introduce
simple mass relations among superflavor partners, primarily derived from superflavor, heavy quark, and light-flavor symmetries.
Then, we demonstrate that these mass relations are somewhat broken among real hadrons, indicating the need for improvements to the mass relations.
Under the superflavor symmetry, a DHT is related to an anti-SHB, and a DHB to an anti-HM.
Considering that $T_{cc}^{+}$ has isospin $I=0$, the superflavor partners of $T_{cc}^{+}$, $T_{ccs}^{+}$, and $T_{ccs}^{++}$ are ,respectively, the antibaryons of $\Lambda_{c}^{+}$, $\Xi_{c}^{+}$, and $\Xi_{c}^{0}$.
Similarly, the superflavor partners of DHBs such as $\Xi_{cc}$ and $\Omega_{cc}$ are the HMs, $\bar{D}$ and $\bar{D}_s$, respectively.

We first study the relation between the masses of the DHBs and HMs in terms of the superflavor symmetry, and then extend the analysis to the DHTs and SHBs.
Based on the HQS, we divide a heavy hadron into a heavy object and light-quark cloud which includes the interaction between them.
Therefore, the dynamics of these hadrons are determined by the properties of the light-quark cloud.
As a result, the masses of heavy hadrons treated in the present analysis are expressed as a sum of the mass of heavy objects and the energy of the light-quark cloud in the heavy-quark limit.
Let us first estimate the masses of HMs.
Here, we consider the spin average of the doublet under the HQS.
We note that due to the spin average, the first-order term in $1/m_Q$ expansion that breaks only the heavy-quark spin symmetry does not appear in the mass formulas.
As a result, the spin-averaged mass can be expressed as
\begin{align}
M_{\rm ave}\left(\bar{Q}q\right)&=M\left(\bar{Q}\right)+E\left(q\right) , 
\label{mass of HM}
\end{align}
where $M(\bar{Q})$ ($\bar{Q}=\bar{c},\bar{b}$) is the mass of antiheavy quark and $E(q)$ ($q=u,d,s$) denotes the contribution from the light-quark cloud.
From Eq.~\eqref{mass of HM}, we obtain the meson mass difference between flavor partners as 
\begin{equation}
M_{\rm ave}(\bar{Q}s) - M_{\rm ave}(\bar{Q}n) = E(s) - E(n) \ , 
\label{HM relation}
\end{equation}
where $n = u, d$.

In the DHBs, the heavy diquark $QQ$ takes the color $\bar{3}$ representation, so that the anti-HMs and the DHBs include the common light-quark cloud in the heavy-quark limit.
Thus, the spin-averaged mass of the doublet members of DHBs is expressed in a similar formula as for the HMs:
\begin{align}
M_{\rm ave}\left(Q Qq \right)&=M\left(QQ \right)+E\left(
q\right) \ . 
\label{mass of DHB}
\end{align}
We stress that the term $E\left( q \right)$ is common in Eqs.~\eqref{mass of HM} and \eqref{mass of DHB}.
The mass difference between flavor partners is $E(\bar{q}_{1})-E(\bar{q}_{2})$, 
where $q_{1},q_{2}=u,d,s$. 
This leads to the following mass relation~\cite{Ma:2017nik}:
\begin{align}
& M_{\rm ave}\left(QQ q_{1}\right)-M_{\rm ave}\left(QQ q_{2}\right) \notag\\
& \quad =M_{\rm ave}
\left(\bar{Q}q_{1}\right)-
M_{\rm ave}\left(\bar{Q}q_{2}\right) \ . 
\label{massrelation1}
\end{align}
This implies that the mass differences between flavor partners are the same in the superflavor partners.

Next, we consider the masses of anti-SHBs and DHTs.
By a similar argument as above, the  mass of an anti-SHB ($\bar{Q}\bar{q}\bar{q}$) is expressed as
\begin{align}
%M_{\rm ave}
M
\left(\bar{Q}\bar{q}_{1}\bar{q}_{2}\right)&=M\left(\bar{Q}\right)+E\left(\bar{q}_{1}\bar{q}_{2}\right) \ . 
\label{mass of SHB}
\end{align}

As we stated above, we assume that two charm quarks inside $T_{cc}$ form a compact diquark and that the diquark belonging to the color $\bar{3}$ representation is dominated.
Therefore, in the heavy-quark limit, the DHT shares a common light-quark cloud with the anti-SHB according to the superflavor symmetry.
Consequently, the mass of DHT is expressed as
\begin{align}
%M_{\rm ave}
M \left(QQ \bar{q}_{1} \bar{q}_{2} \right) & = M\left(QQ\right)+E\left(\bar{q}_{1} \bar{q}_{2}\right)\ . 
\label{mass of DHT}
\end{align}
From the mass formulas in Eqs.~(\ref{mass of SHB}) and (\ref{mass of DHT}), we obtain
the following mass relation corresponding to the mass relation~\eqref{massrelation1}:
\begin{align}
& %M_{\rm ave}
M
\left(QQ\bar{q}_{1}\bar{q}_{2}\right)-
%M_{\rm ave}
M
\left(QQ\bar{q}_{3}\bar{q}_{4}\right)\notag\\
& \quad =
%M_{\rm ave}
M
\left(\bar{Q}\bar{q}_{1}\bar{q}_{2}\right)-
%M_{\rm ave}
M
\left(\bar{Q}\bar{q}_{3}\bar{q}_{4}\right) \ . 
\label{massrelation2}
\end{align}
where $\bar{q}_{i} = \bar{u},\bar{d},\bar{s}$ ($i=1,2,3,4$). 

We further combine Eqs.~(\ref{mass of HM}), (\ref{mass of DHB}), (\ref{mass of SHB}) and (\ref{mass of DHT}) to derive the following simple mass relation:
\begin{align}
& M\left(QQ\bar{q}_{1}\bar{q}_{2}\right)-M\left(\bar{Q}\bar{q}_{1}\bar{q}_{2}\right) \notag\\
& \quad = M_{\rm ave}\left(QQq\right)-M_{\rm ave}\left(\bar{Q}q\right) \ . 
\label{massrelation3}
\end{align}

Now, we compare the obtained simple mass relations with existing experimental data.
We first note that the relation~(\ref{HM relation}) implies that the mass difference between $D$ and $D_{s}$ is equal to that between $B$ and $B_{s}$, namely
\begin{align}
    M_{\rm ave}(D_s)-M_{\rm ave}(D) = M_{\rm ave}(B_s) - M_{\rm ave}(B) \, ,
\end{align}
because the energy difference between light clouds, $E(s)-E(n)$, is independent of the heavy flavor.
Similarly, we obtain the following mass relation for the masses of SHBs:
\begin{align}
    M(\Xi_c)-M(\Lambda_c) = M(\Xi_b) - M(\Lambda_b) \, .
\end{align}
\begin{table}[tb]
    \centering
    \begin{tabular}{ccc}
    \hline\hline
    Hadrons&Mass [MeV]&Input value [MeV]\\
    \hline \hline
    $T_{cc}^{+}$&3874.74&3875\\
    \hline
    $\Xi_{cc}^{+}$&3623.0&3622\\
    $\Xi_{cc}^{++}$&3621.55&3622\\
    \hline
    $D^{0}$&1864.84&1867\\
    $D^{\pm}$&1869.66&1867\\
    $D^{*0}$&2006.85&2009\\
    $D^{* \pm}$&2010.26&2009\\
    $D_{s}^{\pm}$&1968.35&1968\\
    $D_{s}^{*\pm}$&2112.2&2112\\
    \hline
    $\Lambda_{c}^{+}$&2286.46&2286\\
    $\Xi_{c}^{+}$&2467.71&2469\\
    $\Xi_{c}^{0}$&2470.44&2469\\
    \hline
    $D^{*0}D_{s}^{\pm}$&3975.20&3979\\
    $D^{* \pm}D_{s}^{\pm}$&3978.61&3979\\
    $D^{0}D_{s}^{*\pm}$&3977.0&3980 \\
    $D^{\pm}D_{s}^{*\pm}$&3981.9&3980 \\
    
    \hline
    $B^{\pm}$&5279.34&5280\\
    $B^{0}$&5279.66&5280\\
    $B^{*}$&5324.71&5325\\
    $B_{s}^{0}$&5366.92&5367\\
    $B_{s}^{*}$&5415.4&5415\\

    \hline
    $\Lambda_{b}^{+}$&5619.60&5620\\
    $\Xi_{b}^{-}$&5797.0&5794\\
    $\Xi_{b}^{0}$&5791.9&5794\\
        \hline\hline
    \end{tabular}
    \caption{    \label{masshadrons}
 Experimental values~\cite{LHCb:2021vvq,LHCb:2021auc,ParticleDataGroup:2023} and input values in this study.
 }
\end{table}
However, from the experimental values of masses shown in Table~\ref{masshadrons}, the mass differences are obtained as
\begin{align}
M_{\rm ave}\left(D_{s}\right)-M_{\rm ave}\left(D\right)= \SI{103}{MeV}, \label{massdiff_DsD}
\\
M_{\rm ave}\left(B_{s}\right)-M_{\rm ave}\left(B\right)= \SI{90}{MeV}. \label{massdiff_BsB}
\end{align}
and
\begin{align}
M\left(\Xi_{c}\right)-M\left(\Lambda_c\right)= \SI{183}{MeV}, \label{massdiff_XicLambdac}
\\
M\left(\Xi_{b}\right)-M\left(\Lambda_b\right)= \SI{175}{MeV}. \label{massdiff_XibLambdab}
\end{align}
This discrepancy implies the necessity of considering correction terms that break both the heavy-quark flavor symmetry and the light-flavor symmetry.

Similarly, applying the mass relation~\eqref{massrelation3} to $T_{cc}$, we obtain the following mass relation:
\begin{align}
M
\left( T_{cc}^{+} \right)-
M
\left(\Lambda_{c}^{+}\right) = M_{\rm ave}\left(\Xi_{cc}\right)-M_{\rm ave}\left(D\right) \ .
\label{massrelation7}
\end{align}
Since $\Xi_{cc}^{*}$ has not been experimentally confirmed yet, we estimate its mass using the following mass relation obtained from the superflavor symmetry~\cite{Brambilla:2005yk,Fleming:2005pd}:
\begin{align}
M\left(\Xi_{cc}^{*}\right)-M\left(\Xi_{cc}\right)=\frac{3}{4}\left(M\left(D^{*}\right)-M\left(D\right)\right) \ ,\label{massrelation4}
\end{align} 
leading to $M\left(\Xi_{cc}^{*}\right)=\SI{3728}{MeV}$, and thus $M_{\rm ave}(\Xi_{cc}) = 3693$\,MeV.
Substituting this value into Eq.~\eqref{massrelation7}, we get $M\left( T_{cc}^{+} \right)=3970 \: \rm{MeV}$. This value is clearly different from the experimental value, $3875 \: \rm{MeV}$, which motivates us to consider the mixing between a state with a heavy diquark in the color $\bar{3}$ representation and one in the color $6$ representation.

\section{IMPROVED MASS RELATIONS} \label{sec:IMPROVED-MASS-RELATION}

In this section, we construct effective Lagrangian terms for solving the problems raised in the previous section for heavy mesons and the $T_{cc}$. 
As we stated in the previous section, we need to include the terms that simultaneously break the heavy-quark flavor symmetry and SU(3)-flavor symmetry for light quarks to cure the problem of masses of heavy mesons and SHBs.
For the problem of the mass of $T_{cc}$, we include a term leading to the mixing between the states constructed from the heavy diquark in the color $\bar{3}$ representation and the one in the color $6$ representation. 
In addition, we include terms which break the heavy-quark spin symmetry for the HQS doublet.

At first, we define the effective DHB($QQq$) fields with quantum numbers $J^{P}=\frac{1}{2}^{+}$ and $J^{P}=\frac{3}{2}^{+}$ in the heavy-quark limit by combining  the heavy diquark with $J^{P}_{\rm heavy} =1^{+}$ to the light-quark cloud with $J^P_{\rm light} = \frac{1}{2}^{+}$. 
This doubly heavy baryon field ${\mathcal B}$ is expressed as
\begin{align}
\left[ {\mathcal B} \right]_{hh'l} = \left[ P_{+} \gamma_{\mu} C P_{+}^T \right]_{hh'} \psi^{\mu}_l \ , 
\end{align}
where $h$ and $h'$ are spinor indices for heavy-quarks and $l$ is the spinor index for the light quark cloud,
$C= i \gamma_2 \gamma_0 $ is the charge conjugation matrix, and $\psi^\mu$ is the field for the heavy quark spin doublet with $J^P = (1/2^+,3/2^+)$.
We should note that the field ${\mathcal B}$ carries the index to specify the heavy quark flavor $Q$.
But here and henceforth, we omit the index to avoid too many indices for one field.
The projection operator for heavy quark $P_+$ is defined as
$P_{+}=(1+v^\mu\gamma_\mu)/2$, where $v^\mu$ is the velocity of the heavy-diquark.
The $\psi^\mu$ field  satisfies the following constraint:
\begin{align}
& v^\nu \gamma_\nu \, \psi^\mu = \psi^\mu \ .
\end{align}
It is convenient to further decompose the field $\psi^\mu$ into the $J^P = 1/2^+$ field $\psi_{1/2}$ and the $J^P = 3/2^+$ field $\psi_{3/2}^\mu$ as
\begin{equation}
\psi^\mu = \frac{1}{\sqrt{3}} \sigma^{\mu\nu} v_\nu \, \psi_{1/2} + \psi^\mu_{3/2} \ ,
\end{equation}
where
\begin{equation}
\sigma^{\mu\nu} = \frac{i}{2} \left[\gamma^\mu \,,\, \gamma^\nu \right] \ ,
\end{equation}
and the spin-$3/2$ Rarita-Schwinger field $\psi^\mu_{3/2}$ satisfies
\begin{align}
%& v^\nu \gamma_\nu \, \psi^\mu_{3/2} = \psi^\mu_{3/2} \ , \notag\\
& v_\mu \psi^\mu_{3/2} = \gamma_\mu \psi^\mu_{3/2} = 0 \ . 
\end{align}

For later use, we define the conjugate field $\bar{\mathcal B}$ for DHB as
\begin{equation}
\left[ \bar{\mathcal B} \right]_{hh'l} = \left[\gamma_0\right]_{hh_1} \left[{\mathcal B}^\dag \right]_{h_1h_2l_1} \left[\gamma_0\right]_{h_2h'} \left[\gamma_0\right]_{l_1l} \ .
\end{equation}
For realizing the superflavor symmetry, we take the effective DHB field $\mathcal{B}$ and the effective anti-HM field $\bar{H} (\sim \bar{Q} q)$ into a unified field $\Psi$ as
\begin{align}
\Psi = \begin{pmatrix} \bar{H} \\ B \end{pmatrix} \ .
\end{align}
The parity transformations are given by
\begin{align}
& \bar{H} \rightarrow \gamma_{0} \bar{H} \gamma_{0},
\\
& \lbrack B \rbrack_{h h^{\prime} l} \rightarrow \lbrack \gamma_{0} \rbrack_{l l_{1} } \lbrack \gamma_{0} \rbrack_{h h_{1}} \lbrack B \rbrack_{h_{1} h_{2} l_{1}} \lbrack \gamma_{0} \rbrack_{h_{2} h^{\prime}} ,
\end{align}
where $h,h^{\prime},h_{1},h_{2}$ are heavy-quark spinor indices and $l,l_{1}$ are spinor indices for a light cloud.
Since the field $\Psi$ belongs to $3$ representation of the $\rm{SU(3)_{f}}$ light flavor, the $\rm{SU(3)_{f}}$ transformation is given by
\begin{align}
\Psi^{i} \rightarrow U ^{i}_{\: j} \Psi^{j}\ ,
\label{SU3 Psi}
\end{align}
where $i,j$ are the light-flavor indices and $U \in \rm{SU(3)_{f}}$.

Next, we define the effective DHT($\sim \bar{Q} \bar{Q} qq$) fields.
As we stated in the previous section,
we include two DHT fields: 
One is constructed from the heavy diquark with the color $\bar{3}$ representation carrying $J_{\rm heavy}^P=1^+$, which is combined with the light-quark cloud with $J^P_{\rm light}=0^+$ to make the DHT with $J^P = 1^+$.
Another one is constructed from the heavy diquark carrying the color $6$ representation and $J^P_{\rm heavy}=0^+$ combined with $J^P_{\rm light} = 1^+$ to make the one with $J^P = 1^+$.
The former one is denoted as $T_\mu^{(\bar{3})}$ and the latter as $T_{\mu}^{(6)}$.
These fields are defined as
\begin{align}
\left[ T^{(\bar{3})} \right]_{hh'} = \left[ P_{+} \gamma^{\mu} C P_{+}^{T} \right]_{hh'}\,\phi_{\mu} \ ,
\\
\left[ T_{\mu}^{(6)} \right]_{hh'} = \left[ P_{+} \gamma_{5} C P_{+}^{T} \right]_{hh'} \, \varphi_{\mu} \ ,
\end{align}
where $h$ and $h'$ are spinor indices for heavy quarks,
and the upper indices of $T$, $(\bar{3})$ and $(6)$, represent the color representation of heavy diquarks in DHTs. 
We note that $\phi$ and $\varphi_\mu$ stand for the annihilation operators and the same applies to $T$.
We also note that these fields have the index of a light quark-flavor. 
As we said in the Introduction, we consider heavy diquarks made from two heavy quarks with the same flavor.
The light-quark cloud is made from two anti light quarks in the flavor antisymmetric representation.
Then, the fields $T$ belong to $3$ representation of the $\rm{SU(3)_{f}}$ light-flavor symmetry.

Based on the superflavor symmetry, the field $T_{\mu}^{(3)}$ and the effective anti-SHB field $\bar{S} (\sim \bar{Q} \bar{q} \bar{q})$ are arranged into a unified field $\Phi$ as
\begin{align}
\Phi = \begin{pmatrix} \bar{S} \\ T^{(\bar{3})}\end{pmatrix} \ .
\end{align}
The parity transformations are given by
\begin{align}
\bar{S} &\rightarrow \bar{S} \gamma_{0} \ ,
\\
\lbrack T \rbrack_{h h^{\prime}} &\rightarrow \lbrack \gamma_{0} \rbrack_{h h_{1}} \lbrack T \rbrack_{h_{1} h_{2}} \lbrack \gamma_{0} \rbrack_{h_{2} h^{\prime}} \ .
\end{align}
The conjugate fields are defined as
\begin{align}
& S = \gamma_{0} \bar{S}^{\dagger} ,\\
& \lbrack \bar{T} \rbrack_{h h^{\prime}} = \lbrack \gamma_{0} \rbrack_{h h_{1}} \lbrack T^{\dagger} \rbrack_{h_{1} h_{2}} \lbrack \gamma_{0} \rbrack_{h_{2} h^{\prime}} \ .
\end{align}
The $\rm{SU(3)_{f}}$ transformation is 
given by
\begin{align}
\bar{\Phi}_{ij} \rightarrow \left( U^{\ast} \right)_{i}^{\: k} \bar{\Phi}_{kl} \left(U^{\dagger} \right)_{\: j}^{l}
\label{SU3 Phi}
\end{align}

Let us construct the effective Lagrangian for the DHT, SHB, DHB and HM.
As we said in the previous section, we need to include the terms that break simultaneously the heavy-quark flavor symmetry and the light-flavor symmetry.
We first note that the terms that break the heavy-quark flavor symmetry are inversely proportional to the heavy-quark mass, i.e., $\propto 1/m_Q$. 
We also include the term for generating the mixing between $\bar{T}^{(3)}_\mu$ and $\bar{T}^{(\bar{6})}_\mu$, which must include $1/m_Q$ since the mixing vanishes at the heavy-quark limit.
For the light-flavor symmetry breaking, we use the spurion field corresponding to the light-quark mass matrix ${\mathcal M}$ as
\begin{equation}
{\mathcal M} = \begin{pmatrix}
    m_u & & \\ & m_d & \\ & & m_s \\
\end{pmatrix} \ ,
\end{equation}
where $m_u$, $m_d$ and $m_s$ are the current quark masses of $u$, $d$ and $s$ quarks, respectively.
This ${\mathcal M}$ transforms as 
\begin{equation}
\left( {\mathcal M} \right)^i{}_j\to \left( U \right)^i{}_k \, \left( {\mathcal M} \right)^k{}_\ell \, \left( U^\dag \right)^\ell{}_j \ .
\end{equation}

Now, the kinetic and mass terms invariant under the heavy-quark symmetry and SU(3) light-flavor symmetry are given by
\begin{align}
{\mathcal L}_{0} = & -\mbox{tr} \left[ \bar{\Psi} i v \cdot \partial \, \Psi - \Lambda_{\Psi} \, \bar{\Psi} \Psi \right] \notag\\
& {} - \mbox{Tr} \left[ \bar{\Phi} i v \cdot \partial \,  \Phi -   \Lambda_{\Phi} \, \bar{\Phi} \Phi \right] \ ,
\end{align}
where $\Lambda_\Psi$ and $\Lambda_\Phi$ are constants with mass dimension one and tr indicates that the traces in spinor space and heavy-spin space are taken, while Tr implies that the traces in the light-flavor space, in addition to the spinor space and heavy-spin space, are taken.
In the following, we explicitly write the indices for light flavor while we omit the indices for heavy-quark flavor and spins of heavy quarks and light quarks.

A possible term that breaks the SU(3) light-flavor symmetry for the $\Psi$ field is written as 
\begin{align}
{\mathcal L}_{\Psi{\rm-br}} = & + c_\Psi \, \mbox{tr} \, \left[ \bar{\Psi}_i \left( \mathcal{M} \right)^i{}_j \Psi^j\right] \ ,
\end{align}
where $c_\Psi$ is a constant with mass dimension zero.

By noting that two light quarks in the $\Phi$ field are antisymmetric in flavor, the light-flavor breaking term is expressed as
\begin{align}
{\mathcal L}_{\Phi{\rm-br}} = & + c_\Phi\, \mbox{tr} \, \left[ \bar{\Phi}_{ij} \left( \mathcal{M} \right)^j{}_k \Phi^{k i} \right] \  \label{lagrangian phi br},
\end{align}
where $c_\Phi$ is a constant with mass dimension zero.

We next construct possible terms that break both the SU(3) light-flavor symmetry and the heavy-quark flavor symmetry, as well as the heavy quark spin symmetry.
We note that the terms also break the superflavor symmetry, so that the terms should be written separately for superflavor partners.
The terms for $\bar{H}$ (anti-HM) and $B$ (DHB) included in $\Psi$ are expressed as
\begin{align}
& {\mathcal L}_{\rm HB-br} = \notag \\
& -  \frac{\Lambda_{\rm f}}{m_Q} \mbox{tr}\left[ H_i \left( \mathcal{M} \right)^i{}_j \bar{H^j} \right] \notag \\
& {} - \frac{\Lambda_{\rm f}^{\prime}}{2 m_Q} \mbox{tr}\left[ \bar{\mathcal{B}}_i \left( \mathcal{M} \right)^i{}_j \mathcal{B}^j \right] \notag \\
& - \frac{\Lambda_{\sigma}^{2}}{8m_Q} \mbox{tr}\left[ H \left( \sigma^{\mu \nu}_{heavy} \right)^T \bar{H} \sigma^{light}_{\mu \nu} \right] \notag \\
& - \frac{\Lambda_{\sigma \rm f}}{8m_Q} \mbox{tr}\left[ H_i \left( \sigma^{\mu \nu}_{heavy} \right)^T \left( \mathcal{M} \right)^i{}_j \bar{H}^j \sigma^{light}_{\mu \nu} \right] \notag \\
& - \frac{\left(\Lambda_{\sigma}^{\prime}\right)^{2}}{8m_Q} \left[ \left[ \bar{\mathcal{B}} \right]_{h_1 h_2 l_1} \left( \sigma^{\mu \nu}_{heavy} \right)_{h_2 h_3} \left[ \mathcal{B} \right]_{h_3 h_1 l_2} \left( \sigma^{light}_{\mu \nu} \right)_{l_2 l_1} \right] \notag \\
& - \frac{ \Lambda_{\sigma \rm f}^{\prime}}{8m_Q} \Bigg[ \left[ \bar{\mathcal{B}}_i \right]_{h_1 h_2 l_1} \left( \sigma^{\mu \nu}_{heavy} \right)_{h_2 h_3} \left( \mathcal{M} \right)^i{}_j \notag\\
& \qquad\qquad\quad \left[ \mathcal{B}^j \right]_{h_3 h_2 l_2} \left( \sigma^{light}_{\mu \nu} \right)_{l_2 l_1} \Bigg] \ , 
\end{align}
where $\Lambda_{\rm f}$, $\Lambda_{\rm f}^{\prime}$, $\Lambda_{\sigma}$, $\Lambda_{\sigma \rm f}$, $\Lambda^{\prime}_{\sigma}$, $\Lambda^{\prime}_{\sigma \rm f}$ are  constants with mass dimension one.
We put an extra factor $1/2$ in the second term since the DHB includes two heavy quarks. 
We note that, if we set $\Lambda_{\sigma} = \Lambda^{\prime}_{\sigma}$ and $\Lambda_{\sigma \rm f} = \Lambda^{\prime}_{\sigma \rm f}$, these two terms reproduce the mass relation~(\ref{massrelation4}).

Similarly, possible breaking terms for $\bar{S}$ (anti-SHB) and $T$ (DHT) in $\Phi$ field are expressed as
\begin{align}
& {\mathcal L}_{\rm ST-br} = - \frac{\Lambda_{\rm{ff}}}{m_Q} \, \mbox{tr} \left[ \bar{S}_{ij}  \left( \mathcal{M} \right)^j{}_k S^{ki} \right]  \notag\\
& \quad {} - \frac{\Lambda_{\rm{ff}}^{\prime}}{2m_Q} \, \mbox{tr} \left[ \left( T^{(\bar{3})} \right)_{ij} \left( \mathcal{M} \right)^j{}_{k} \left( \bar{T}^{(\bar{3})} \right)^{ki} \right]  \ ,
\end{align}
where $\Lambda_{\rm{ff}}$ and $\Lambda_{\rm{ff}}^{\prime}$ are constants with mass dimension one.

Finally, we consider a DHT field $T_\mu^{(6)}$  constructed from the heavy diquark with color ${\mathbf{6}}$ representation.
The kinetic and mass terms are written as
\begin{equation}
{\mathcal L}_{T^{(6)}} = - \mbox{Tr} \, \left[ \bar{T}_\mu^{(6)} \left( i v \cdot \partial - \Lambda_6 \right) T^{(6) \mu} \right] \ ,
\end{equation}
where $\Lambda_6$ is a constant with mass dimension one.
The term for the mixing between two DHT fields $T_\mu^{(3)}$ and $T_\mu^{(6)}$ is expressed as
\begin{align}
{\mathcal L}_{\rm{mix}} = & - \frac{\Lambda_{\rm{mix}}^{2}}{2m_Q} \mbox{Tr} \Bigg[ \left( \bar{T}_{\mu}^{(6)} \right) \gamma_{5} \gamma^{\mu} \left( T^{(\bar{3})} \right) + h.c. \Bigg] \ . 
\end{align}

From the above Lagrangian terms, the masses of HM, DHB and SHB are modified from Eqs.~\eqref{mass of HM}, \eqref{mass of DHB} and \eqref{mass of SHB} as
\begin{align}
M\left(\bar{Q}q\right) = &\, M\left(\bar{Q}\right)+E\left(q\right) \notag\\
& {} + \frac{m_{q}}{m_{Q}} \Lambda_{\rm f} - \frac{3}{4} \frac{\Lambda^{2}_{\sigma}}{m_Q} - \frac{3}{4} \frac{m_q}{m_Q} \Lambda_{\sigma \rm f} , \label{mass of ground HM} \\
M^{*}\left(\bar{Q}q\right) = &\, M\left(\bar{Q}\right)+E\left(q\right) \notag\\
& {} + \frac{m_{q}}{m_{Q}} \Lambda_{\rm f} + \frac{1}{4} \frac{\Lambda^{2}_{\sigma \rm f}}{m_Q} + \frac{1}{4} \frac{m_q}{m_Q} \Lambda_{\sigma \rm f} , \label{mass of excited HM} \\
M\left(QQq\right) = & M\left(QQ\right)+E\left(q\right) \notag\\
& {} + \frac{m_{q}}{2 m_{Q}} \Lambda_{\rm f}^{\prime} - \frac{1}{2} \frac{\left( \Lambda^{\prime}_{\sigma} \right)^{2} }{m_Q} - \frac{1}{2} \frac{m_q}{m_Q} \Lambda^{\prime}_{\sigma \rm f} , \label{mass of ground DHB} \\
M^{*}\left(QQq\right) = &\, M\left(QQ\right)+E\left(q\right) \notag\\
& {} + \frac{m_{q}}{2 m_{Q}} \Lambda_{\rm f}^{\prime} + \frac{1}{4} \frac{\left( \Lambda^{\prime}_{\sigma} \right)^{2} }{m_Q} + \frac{1}{4} \frac{m_q}{m_Q} \Lambda^{\prime}_{\sigma \rm f} , \label{mass of excited DHB} \\
%M_{\rm ave}
M\left(\bar{Q}\bar{q_{1}}\bar{q_{2}}\right) = & M\left(\bar{Q}\right)+E\left(\bar{q_{1}}\bar{q_{2}}\right) 
+ \frac{m_{q_{1}} + m_{q_{2}}}{m_{Q}} \Lambda_{\rm f} , \label{mass of SHB 2} 
\end{align}
where
\begin{align}
E(q) & = \Lambda_\Psi + c_\Psi m_q \ , \notag\\
E(\bar{q}_1 \bar{q}_2 ) & = \Lambda_\Phi + c_\Phi \left( m_{q_1} + m_{q_2} \right) \ .
\end{align}
The square mass matrix for two DHT fields is expressed as
\begin{equation}
\begin{pmatrix}
    M_3^2 & \frac{\Lambda_{\rm{mix}}^2}{2 m_Q}\sqrt{M_6 M_3} \\
    \frac{\Lambda_{\rm{mix}}^2}{2 m_Q}\sqrt{M_6 M_3} & M_6^2 \\
\end{pmatrix} \ ,
\end{equation}
where
\begin{equation}
M_3 = M\left(QQ\right)+E\left(\bar{q_{1}}\bar{q_{2}}\right) + \frac{m_{q_{1}} + m_{q_{2}}}{2 m_{Q}} \Lambda_{\rm{ff}}^{\prime} \ ,
\end{equation}
and $M_6$ is the mass of the $T_\mu^{(6)}$ field before mixing.
By diagonalizing this matrix up to $1/m_Q$ order, the mass of the lightest DHT is obtained as 
\begin{align}
& 
%M_{\rm ave}
M\left(QQ\bar{q_{1}}\bar{q_{2}}\right) = M\left(QQ\right)+E\left(\bar{q_{1}}\bar{q_{2}}\right) \notag \\ 
& \quad {} + \frac{m_{q_{1}} + m_{q_{2}}}{2 m_{Q}} \Lambda_{\rm{ff}}^{\prime} - \frac{M_6}{2(M_6^2-M_3^2)} \left( \frac{\Lambda_{\rm{mix}}^{2}}{2 m_{Q} } \right)^2 \ . \label{mass of DHT 2}
\end{align}
By combining the above mass formulas, the simple mass relations~\eqref{massrelation1}, \eqref{massrelation2} and \eqref{massrelation3} are modified as:
\begin{align}
&M_{\rm ave}
\left(Q Q s\right)-
M_{\rm ave}
\left(QQn\right)-\frac{m_{s}-m_{n}}{2m_{Q}} \Lambda_{\rm f}^{\prime} \notag\\ 
&=
M_{\rm ave}
\left(\bar{Q} s \right)-
M_{\rm ave}
\left(\bar{Q} n \right)-\frac{m_{s}-m_{n}}{m_{Q}} \Lambda_{\rm{f}},
\label{improved mass relation 1} \\
&M
\left(QQ \bar{s}\bar{n} \right)-
M
\left(QQ \bar{u}\bar{d} \right)-\frac{m_{s}-m_{n}}{2m_{Q}} \Lambda_{\rm{ff}}^{\prime} \notag\\
&=M
\left(\bar{Q} \bar{s} \bar{n} \right)-
M
\left(\bar{Q} \bar{u} \bar{d}\right)-\frac{m_{s}-m_{n}}{m_{Q}} \Lambda_{\rm{ff}},\\
&M\left(QQ \bar{u}\bar{d}\right)-M\left(\bar{Q} \bar{u} \bar{d}\right) \notag\\
&\ \ {}+\frac{M_{6}}{2 \left(M_{6}^{\: 2}-M_{3}^{\: 2}\right)}\left(\frac{\Lambda_{\rm{mix}}^{2}}{2m_{Q}}\right)^{2}-\frac{m_{n}}{2m_{Q}}\left(\Lambda_{\rm{ff}}^{\prime}-2 \Lambda_{\rm{ff}} \right) \notag\\
& =M_{\rm ave}\left(QQn\right)-M_{\rm ave}\left(\bar{Q}n\right) -\frac{m_{n}}{2m_{Q}}\left(\Lambda_{\rm{f}}^{\prime}-2 \Lambda_{\rm{f}} \right) \ ,
\label{improved relation}
\end{align}
\begin{comment}
\begin{align}
&M\left(\bar{Q} \bar{Q} \bar{s}\right)-M\left(\bar{Q}\bar{Q}\bar{n}\right)-\frac{m_{s}-m_{n}}{2m_{Q}} \Lambda_{\rm f}^{\prime} \notag\\ 
&=M\left(Q\bar{s}\right)-M\left(Q\bar{n}\right)-\frac{m_{s}-m_{n}}{m_{Q}} \Lambda_{\rm f},
\label{improved mass relation 1} \\
&M\left(\bar{Q}\bar{Q}sn\right)-M\left(\bar{Q}\bar{Q}ud\right)-\frac{m_{s}-m_{n}}{2m_{Q}} \Lambda_{\rm{ff}}^{\prime} \notag\\
&=M\left(Qsn\right)-M\left(Qud\right)-\frac{m_{s}-m_{n}}{m_{Q}} \Lambda_{\rm{ff}},\\
&M_{\rm ave}\left(\bar{Q}\bar{Q}ud\right)-M_{\rm ave}\left(Qud\right) \notag\\
&+\frac{M_{6}}{M_{6}^{\: 2}-M_{3}^{\: 2}}\left(\frac{\Lambda_{mix}^{2}}{2m_{Q}}\right)^{2}-\frac{m_{n}}{m_{Q}}\left(\Lambda_{\rm{ff}}^{\prime}-2 \Lambda_{\rm{ff}} \right) \notag\\
& =M_{\rm ave}\left(\bar{Q}\bar{Q}\bar{n}\right)-M_{\rm ave}\left(Q\bar{n}\right) -\frac{m_{n}}{2m_{Q}}\left(\Lambda_{\rm{ff}}^{\prime}-2 \Lambda_{\rm{ff}} \right) \ ,
\label{improved relation}
\end{align}
\end{comment}
where $n=u,d$ and $m_n = (m_u + m_d)/2$ is the isospin-averaged mass of up and down quarks.
We note that other mass relations such as the ones between non-spin-averaged masses are obtained by recombining Eqs.~\eqref{mass of ground HM}-\eqref{mass of SHB 2}, and \eqref{mass of DHT 2}.
Furthermore, from Eqs.~\eqref{mass of ground HM}, \eqref{mass of excited HM} and \eqref{mass of SHB 2}, we obtain the following mass relation:
\begin{align}
& \left[ M_{\rm ave}(D_s) - M_{\rm ave}(D) \right]
- \left[ M_{\rm ave}(B_s) - M_{\rm ave}(B) \right] \notag\\
& \quad = \left( m_s - m_n \right) \left( \frac{1}{m_c} -\frac{1}{m_b} \right) \Lambda_{\rm{f}} \ , \\
& \left[ M(\Xi_c) - M(\Lambda_c) \right]
- \left[ M(\Xi_b) - M(\Lambda_b) \right] \notag\\
& \quad = \left( m_s - m_n \right) \left( \frac{1}{m_c} -\frac{1}{m_b} \right) \Lambda_{\rm{ff}} \
\end{align}
where $n=u,d$.
\begin{table}[tb]
    \centering
    \begin{tabular}{cc}
    \hline\hline
    Quarks&Mass [MeV]\\
    \hline \hline
    $u,d$&3.415\\
    $s$&93.40\\
    $c$&1270\\
    $b$&4180\\
        \hline\hline
    \end{tabular}
\caption{    \label{quarkmass}
 Mass of quarks~\cite{ParticleDataGroup:2023} used as inputs.
}
\end{table}
Using the masses shown in Table~\ref{masshadrons} together with the current quark masses shown in Table~\ref{quarkmass}, 
we determine the value of parameters as
\begin{align}
\Lambda_{\rm f} = 272.2 \pm 70.9 \, \mbox{MeV} \ , \label{lambda_f} \\
\Lambda_{\rm{ff}} = 157.4 \pm 70.9 \,\mbox{MeV} \ , \label{lambda_ff} \\
\Lambda_{\sigma} = 423.5 \pm 25.1 \, \mbox{MeV} \ , \label{lambda_s} \\
\Lambda_{\sigma \rm f} = 36.70 \pm 70.87 \,\mbox{MeV} \ . \label{lambda_sf} 
\end{align}
The error bars are estimated by 
\begin{align}
\delta \Lambda_{\rm f} & =\delta \Lambda_{\rm{ff}}=\delta \Lambda_{\sigma \rm f}=\frac{\Lambda_{\rm{QCD}}^{2}}{m_{Q}} , \, \notag \\
\delta \Lambda_{\sigma} & =\frac{\Lambda_{\rm{QCD}}^{3}}{2 \Lambda_{\sigma} m_{Q}} ,
\end{align}
where $\Lambda_{\rm QCD}$ is the typical scale of QCD and we set $\Lambda_{\rm QCD}=300 \, \mbox{MeV}$ in this paper. The parameters $\Lambda_{\sigma}$ and $\Lambda_{\sigma \rm f}$ are determined to fit the hyperfine splittings between charmed mesons, and we confirmed that the hyperfine splittings between bottom mesons are within error bars. The other parameters $\Lambda_{\rm f}^{\prime}$ and $\Lambda_{\rm{ff}}^{\prime}$ cannot be determined due to lack of input hadron masses. We therefore assume that they are of the same order as $\Lambda_{\rm f}$ and $\Lambda_{\rm{ff}}$. Specifically, we set
\begin{align}
\Lambda_{\rm f}^{\prime} = 0 \pm 378.4 \, \mbox{MeV} \ , \label{lambda_f'} \\
\Lambda_{\rm{ff}}^{\prime} = 0 \pm 263.6 \,\mbox{MeV} \ , \label{lambda_ff'} \\
\Lambda_{\sigma}^{\prime} = 0 + 448.6 \, \mbox{MeV} \ , \label{lambda_s'} \\
\Lambda_{\sigma \rm f}^{\prime} = 0 + 107.9 \,\mbox{MeV} \ . \label{lambda_sf'} 
\end{align}
The conditions $\Lambda_{\sigma}^{\prime}$ and $\Lambda_{\sigma \rm f}^{\prime}$ are required to be non-negative, based on the assumption that the $J= \frac{3}{2}$ state does not have a smaller mass than the $J= \frac{1}{2}$ state.
By applying the improved mass relation corresponding to Eq.~\eqref{massrelation2} to $T_{cc}^{+}$, we derive
\begin{align}
& M
\left(T_{ccs}\right) - 
M
\left(T_{cc}^{+}\right) - \frac{m_{s} - m_{n}}{2 m_{c}} \Lambda_{\rm{ff}}^{\prime} \notag \\
& \quad = M
\left(\Xi_{c}\right) - 
M
\left(\Lambda_{c}^{+}\right) - \frac{m_s - m_{n}}{m_{c}} \Lambda_{\rm{ff}} \ , \label{massrelation5}
\end{align}
and we obtain the mass of $T_{ccs}$ as
\begin{align}
M\left(T_{ccs}\right)=4047 \pm 11 \: \rm{MeV} . \label{mass of tccs}
\end{align}
We note that the above value corresponds to the isospin-averaged masses of $T_{ccs}^{+}$ and $T_{ccs}^{++}$. 
\begin{table}[tb]
    \centering
    \begin{tabular}{ccc}
    \hline\hline
    Hadrons&Mass [MeV]& \\
    \hline \hline
    \multirow{2}{*}{$T_{ccs}$}&$4047 \pm 11$&Our result\\
    &$4106$&NQM~\cite{Karliner:2021wju}\\
    $T_{ccs}^{+}$&$3969 \pm 8$&Lattice QCD~\cite{Junnarkar:2018twb}\\
        \hline\hline
    \end{tabular}
    \caption{    \label{results of Tccs}
 Theoretical predictions for the masses of $T_{ccs}$. NQM imply a nonrelativistic quark model.}
\end{table}
If this result agrees with future experimental data, it means that the color antitriplet state of the heavy diquark is dominant in $T_{cc}^+$ and $T_{ccs}$. 
We compare our result with other results in Table~\ref{results of Tccs}.

From the mass relation in Eq.~(\ref{improved mass relation 1}), we further obtain
\begin{align}
M\left(\Omega_{cc}\right) &- M\left(\Xi_{cc}\right) - \frac{m_{s} - m_{n}}{2 m_{c}} \left( \Lambda_{\rm f}^{\prime} - \Lambda_{\sigma \rm f}^{\prime} \right) \notag \\
& = M_{\rm ave}\left(D_{s}\right) - M_{\rm ave}\left(D \right) - \frac{m_{s} - m_{n}}{m_{c}} \Lambda_{\rm f} , \label{massrelation6}
\end{align}
where we used spin-averaged masses for $D_{s}$ and $D$ to reduce the ambiguity from the correction of the $\Lambda_{\sigma \rm f}$ term.
From this relation, the mass of $\Omega_{cc}$, which has not been experimentally reported so far, is also predicted as 
\begin{align}
M\left(\Omega_{cc}\right)= 3706^{+14}_{-15} \: \rm{MeV} . \label{mass of omegacc}
\end{align}
This result serves as another indicator for assessing whether our mass relations are correct.
If this result does not match future experimental results, one possible reason might be that $\Lambda_{\rm f}^{\prime}$ is larger than $\Lambda_{\rm f}$.
Additionally, the second-order effects of $1/m_Q$ might also contribute.
We show comparison with other results in Table~\ref{Omega_cc masses}.
\begin{table}[tb]
    \centering
    \begin{tabular}{ccc}
    \hline\hline
    Hadrons &Mass [MeV]&\\
        \hline\hline    
    \multirow{8}{*}{$\Omega_{cc}$} &$3706^{+14}_{-15}$ & Our result\\
    &$3766 \pm 2$& NQM~\cite{Ortiz-Pacheco:2023kjn}\\
    &3715 & RQM~\cite{Lu:2017meb}\\
    &3778 & RQM~\cite{Ebert:2002ig}\\
    &3590 & NQM~\cite{Gershtein:2000nx}\\
    &3815 & NQM~\cite{Roberts:2007ni}\\
    &$3733 \pm 13$ & Lattice QCD~\cite{PhysRevD.102.054513}\\
    &$3704 \pm 17$ & Lattice QCD~\cite{PACS-CS:2013vie}\\
        \hline\hline
    \end{tabular}
    \caption{Several theoretical predictions of the mass of $\Omega_{cc}$. NQM and RQM imply nonrelativistic and relativistic quark models, respectively.}
    \label{Omega_cc masses}
\end{table}
Since hadrons containing two $b$ quarks have not yet been experimentally discovered, we cannot predict in the bottom sector. However, we can predict mass differences as
\begin{align}
M\left(T_{bbs}\right) &- M\left(T_{bb}\right) = 172 \pm 3 \: \rm{MeV} , , \label{mass diff DHT b}
\end{align}
and
\begin{align}
M\left(\Omega_{bb}\right) &- M\left(\Xi_{bb}\right) = 84^{+4}_{-5} \: \rm{MeV} . \label{mass diff DHB b}
\end{align}
We compare the above results with the results obtained in some lattice analyses in Table~\ref{mass differences of bottom hadrons}. 
This shows that our results are consistent with lattice QCD results except for those in Refs.~\cite{Meinel:2022lzo,Wagner:2022bff}.
\begin{table}[tb]
    \centering
    \begin{tabular}{ccc}
    \hline\hline
    Hadrons &Mass [MeV]&\\
        \hline\hline    
    \multirow{4}{*}{$M\left(T_{bbs}\right)-M\left(T_{bb}\right)$} &$172 \pm 3$ & Our result\\
    &$178 \pm 12 \pm 4$&Lattice QCD~\cite{PhysRevLett.118.142001}\\
    &$143 \pm 47$ &Lattice QCD~\cite{PhysRevD.99.034507}\\
    &$104 \pm 23 \pm 10$ &Lattice QCD~\cite{Meinel:2022lzo,Wagner:2022bff}\\
        \hline
    \multirow{4}{*}{$M_{ave}\left( \Omega_{bb} \right) -M_{ave}\left( \Xi_{bb} \right) $} &$84^{+4}_{-5}$ & Our result\\
    &$96 \pm 13^{+17}_{-21}$&Lattice QCD~\cite{PhysRevD.79.014502}\\
    &$130 \pm 25 \pm 23$ &Lattice QCD~\cite{Brown:2014ena}\\
    &$99 \pm 23$ &Lattice QCD~\cite{Mohanta:2019mxo}\\
        \hline\hline
    \end{tabular}
    \caption{Several theoretical predictions of the mass differences of bottom hadrons. For tetraquarks, we estimated the masses in the lattice QCD by subtracting binding energies from physical thresholds, i.e. $M(T_{bb})=M(B^{*})+M(B)- \Delta E(T_{bb}), \, M(T_{bbs})=M(B^{*})+M(B_{s})- \Delta E(T_{bbs})$, where $\Delta E(T_{bb})$ and $ \Delta E(T_{bbs})$ are the binding energies of $T_{bb}$ and $T_{bbs}$ shown in Refs.\cite{PhysRevLett.118.142001,PhysRevD.99.034507,Meinel:2022lzo,Wagner:2022bff}, 
    respectively.}
    \label{mass differences of bottom hadrons}
\end{table}

\section{WIDTH OF $T_{ccs}$} \label{width}

In this section, we construct the Lagrangian solely from the $\rm{SU(3)}$ light-flavor symmetry and predict the decay width of $T_{ccs}$ using the decay width of $T_{cc}^{+}$ as an input.
Our approach hinges on the fact that both $T_{ccs}$ and $T_{cc}^{+}$ belong to the same light-flavor representation. Notably, this method is independent of both superflavor symmetry and color representation.

\begin{figure*}[tb]
\centering
\begin{tabular}{ccc}
\includegraphics[scale=0.8]{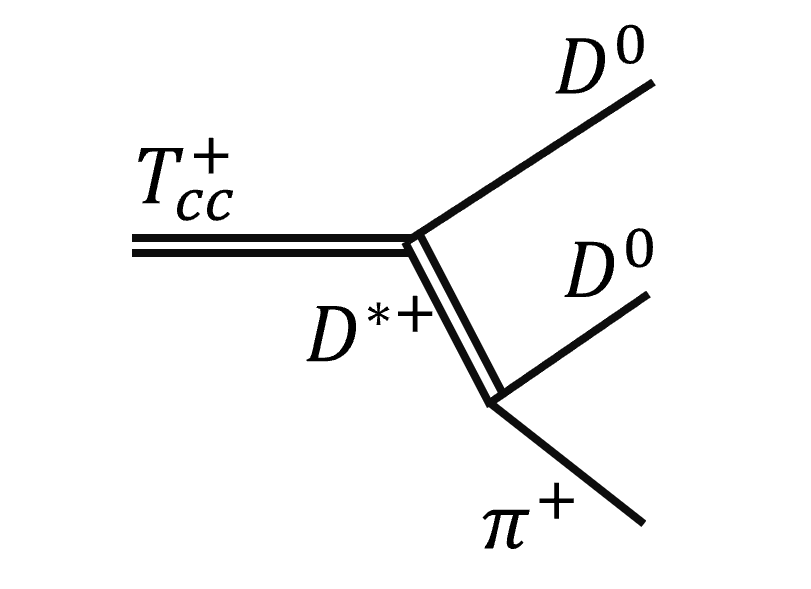}&
\includegraphics[scale=0.8]{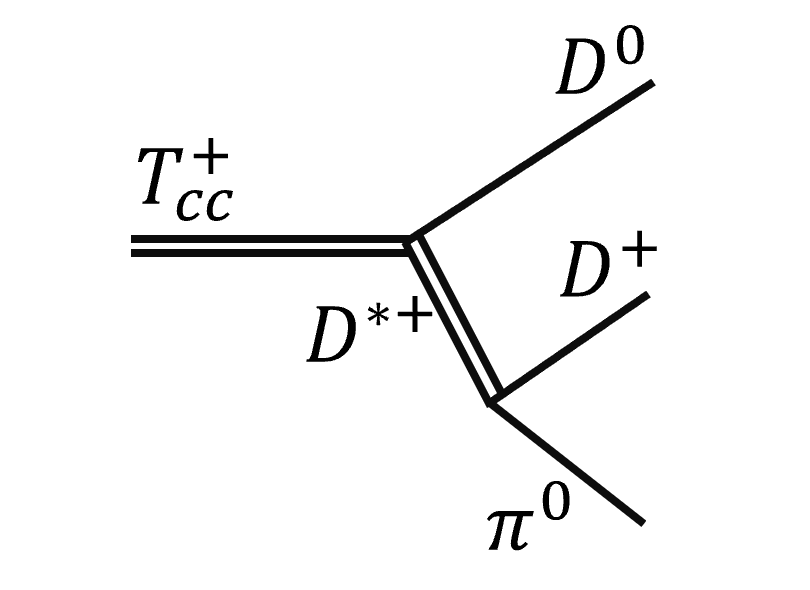}&
\includegraphics[scale=0.8]{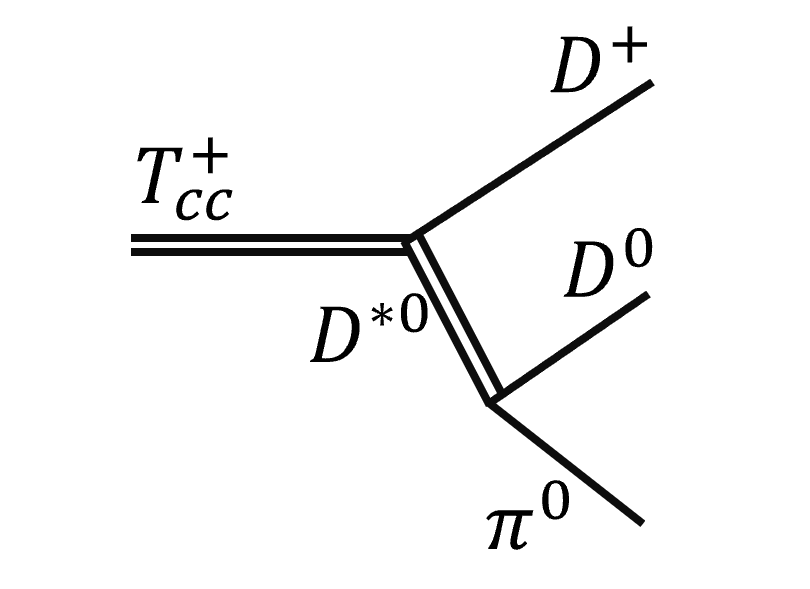}\\
(a)&(b)&(c)
\end{tabular}
\caption{Feynmann diagrams of the $T_{cc}^{+} \to DD\pi$ decays.}
\label{tccdiag}
\end{figure*}

\begin{figure*}[tb]
\centering
\begin{tabular}{cc}
\includegraphics[scale=0.8]{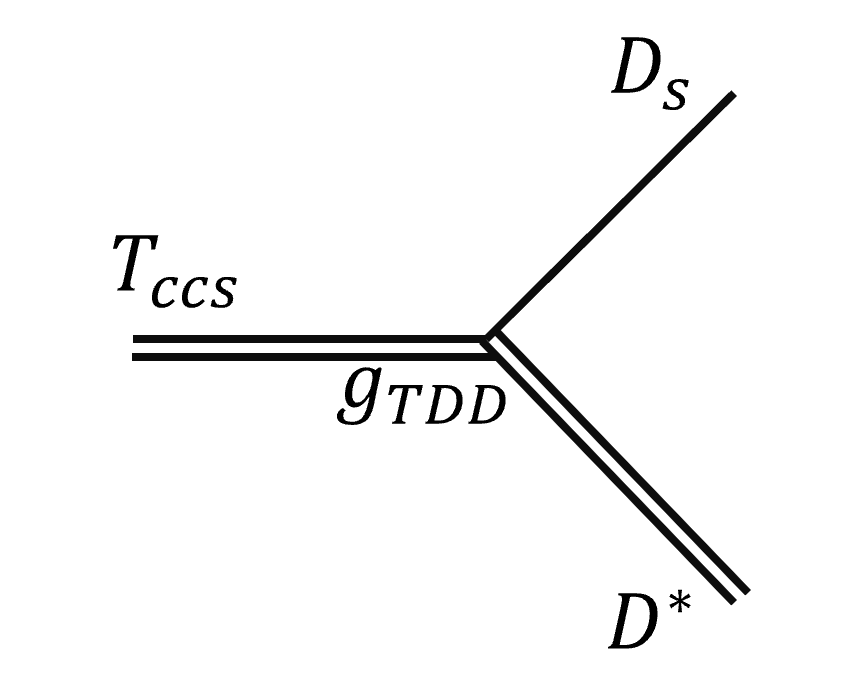}&
\includegraphics[scale=0.8]{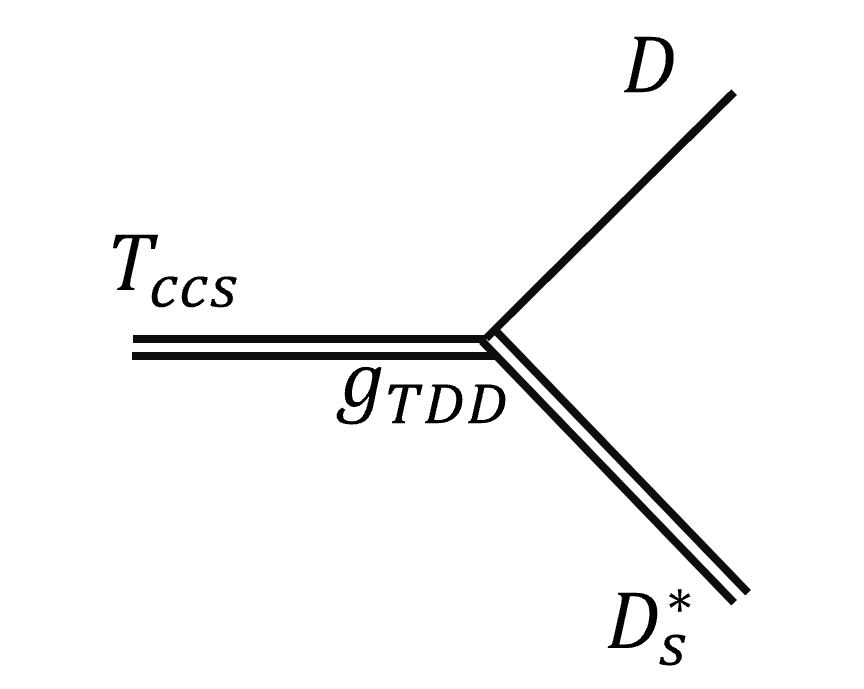}\\
(a)&(b)
\end{tabular}
\caption{Feynmann diagrams of the $T_{ccs} \to DD_s^{*}$ and $D^{*}D_s$ decays.}
\label{tccsdiag}
\end{figure*}

Since $T_{cc}^+$ locates just below the $DD^*$ threshold, we consider the decay process of $T_{cc}^{+} \rightarrow D^{0}D^{0}\pi^{+}$ and $T_{cc}^{+} \rightarrow D^{0}D^{+}\pi^{0}$ with $D^{*}$ as an intermediate state, as depicted in Fig.~\ref{tccdiag}. 
While the decay $T_{cc}^{+} \rightarrow D^{0}D^{+}\pi^{0}$ has not been observed in experiments~\cite{LHCb:2021vvq,LHCb:2021auc}, it is not prohibited kinematically. 
Hence, in this study, we also incorporate Figs.~\ref{tccdiag}(b) and ~\ref{tccdiag}(c). 
For $T_{ccs}$, we consider the decay processes $T_{ccs} \rightarrow DD_{s}^{*}$ and $T_{ccs} \rightarrow D^{\ast} D_{s}$ as shown in Fig.~\ref{tccsdiag}, because the mass of $T_{ccs}$ predicted in the previous section is above the $DD_s^*$ and $D^*D_s$ thresholds. 

Let us construct an effective Lagrangian for the interaction among the tetraquarks and charmed mesons based on just light-flavor symmetry. 
In the following we use relativistic field $T^{\mu}$ for the mass eigenstates of flavor triplet tetraquarks:
\begin{equation}
T^{\mu} = \begin{pmatrix}
    0 & T_{cc}^{+} & T_{ccs}^{+} \\ -T_{cc}^{+} & 0 & T_{ccs}^{++} \\ -T_{ccs}^{+} & -T_{ccs}^{++} & 0 \\
\end{pmatrix} \ \label{massmatrix_light},
\end{equation}
While the fields for charmed mesons with $J^{P}=\left(0^{-},1^{-}\right)$ are represented by relativistic fields $\left(D, D^{*\mu}\right)$:
\begin{equation}
D = \begin{pmatrix}
D^0 \\ D^+ \\ D_s \\
\end{pmatrix} \ , \quad
D^\ast = \begin{pmatrix}
D^{\ast0} \\ D^{\ast+} \\ D_s^\ast \\
\end{pmatrix} \ .
\end{equation}
From Eqs.~(\ref{SU3 Psi}) and (\ref{SU3 Phi}), the 
$\rm{SU(3)}$ light-flavor transformations of these fields are given as 
\begin{align}
& T_{ij}^{\mu} \rightarrow \left( U^{\ast} \right)_{i}^{\: k} T_{kl}^{\mu} \left(U^{\dagger} \right)_{\: j}^{l} , \notag \\
%\: 
& D_{i} \rightarrow D_{j}\left(U^{\dag}\right)^{j}_{\; i} ,\: D_{i}^{*\mu} \rightarrow D_{j}^{*\mu}\left(U^{\dag}\right)^{j}_{\; i} \ .
\label{ftrans}
\end{align}
The Lagrangian for $T_{cc(s)}$ and heavy mesons invariant under this transformation is given by
\begin{align}
\mathcal{L}=g_{TDD} \bar{D}_{\mu}^{\ast i} T_{ij}^{\mu} \bar{D}^{j} \ .
\label{ltdd}
\end{align}

As we said above, we determine the value of $g_{TDD}$ from the decay width of $T_{cc}$, $\SI{48}{keV}$, assuming that the decay is dominated by the process shown in Fig.~\ref{tccdiag}.
By using the value of the $D^\ast D \pi$ coupling constant determined from the decay of the $D^\ast$ meson (see, e.g., Ref~\cite{Manohar:2000dt}), the value of $g_{TDD}$ is calculated as
\begin{align}
g_{TDD}= \left( 4.2 \pm 1.2 \right) \times 10^{3} \: \rm{MeV} \ . 
\label{gtdd}
\end{align}
We estimate the error to be 30\% based on $\frac{ms}{\Lambda_{\rm QCD}}$.
We note that this value is natural when nondimensionalized 
by the mass of the charm quark. 
Now, based on the processes shown in Fig.~\ref{tccsdiag}. the formula of the decay width of $T_{ccs}$ is calculated as
\begin{align}
\Gamma \left(T_{ccs}\right)=g_{TDD}^{2}\frac{\left|P_{1}\right|+\left|P_{2}\right|}{6\pi M \left(T_{ccs}\right)^{2}} \ ,
\end{align}
where $P_1$ and $P_2$ are phase space momenta corresponding to Figs.~\ref{tccsdiag}(a) and ~\ref{tccsdiag}(b), respectively.
By using the mass of $T_{ccs}$ predicted in Eq.~\eqref{mass of tccs} together with the value of $g_{TDD}$ in Eq.~(\ref{gtdd}), the decay width is predicted as
\begin{align}
\Gamma \left(T_{ccs}\right)=42 \pm 24 \: \rm{MeV}  \ .
\end{align}
We consider this decay width to be sufficiently small as to be experimentally observable.

\section{SUMMARY} \label{sec:summary}

In this study, we investigated the mass and decay width
of the doubly heavy tetraquark $T_{ccs}$ from the superflavor and $\rm{SU(3)}$ light-flavor symmetries.
We assumed that doubly charmed tetraquarks are constructed from a color antitriplet $cc$ diquark and thus they are the superflavor partners of the singly heavy baryons.
First, we derived the simple mass relations under heavy quark and superflavor symmetries.
However, we found a discrepancy between predictions of the obtained mass relations and the experimental data.
Then, we constructed an effective Lagrangian based on these symmetries by including correction terms violating simultaneously the heavy-quark symmetry and the light-flavor symmetry. 
From the Lagrangian, we obtained the improved mass relations among heavy mesons, doubly heavy baryons, singly heavy baryons, and doubly heavy tetraquarks. 
Based on the relations, we predicted the mass of unobserved tetraquark $T_{ccs}$ as $M\left(T_{ccs}\right)=4047 \pm 11 \: \rm{MeV}$.
We also predicted the mass of unobserved $\Omega_{cc}$ as $M(\Omega_{cc})=3706 ^{+14}_{-15} \, \rm{MeV} $.

We then constructed an effective Lagrangian term for the decay of $T_{ccs}$ based on SU(3) light-flavor symmetry.
The unknown coupling constant was determined by using the $T_{cc}$ decay data. Incorporating the predicted mass, we derived the decay width of $T_{ccs}$ as $\Gamma \left(T_{ccs}\right)=42 \pm 24 \: \rm{MeV}$.

The obtained masses and widths will be useful to understand the color configuration of DHTs.
If these results agree with future experimental data, it means that the color antitriplet state in $T_{cc}^{+}$ and $T_{ccs}$ is dominant.

The isovector counterpart of the isoscalar $T^{+}_{cc}$ can also exist independently from $T^{+}_{cc}$, and we expect that the analyses can be done separately.
It is interesting to study the isovector state by extending the present analysis, which we leave to the future work.

\section{ACKNOWLEDGMENT} \label{sec:acknowledment}
This work was supported in part by JSPS KAKENHI Grants Nos.~JP20K03927, No. JP23H05439, No. JP24K007045 (M.H,), and No. JP20K14478 (Y.Y,).
%\nocite{*}

\bibliography{main}

\end{document}